\documentclass[twocolumn,9pt]{article} % here we use the article class, rather than elsarticle

\usepackage[square,numbers,sort&compress,comma]{natbib}

\usepackage{amsmath}
\usepackage{amssymb}
\usepackage{caption}
\usepackage{graphicx}
\usepackage{latexsym}
\usepackage{times}
\usepackage[pagewise]{lineno}
\usepackage{hyperref}
\usepackage{soul}

\usepackage[version=4]{mhchem}
\usepackage{xcolor}

\topmargin - 12pt % might need to be set to 0pt for some installations
\renewenvironment{abstract}%
              {% - begin definition
               \small% - select font
               {\bfseries \abstractname}% - select font
               \par% - end a paragraph (skip \parsep)
               \vspace{10pt}% - add vertical space
              }% - complete definition

\renewcommand\abstractname{Abstract}

\newcommand{\nomenclature}% - name of command
              [1]% - number of arguments
              {% - begin definition
               \bgroup% - begin a local group
               \flushleft% - turn on flushleft option
               \small\bf% - select font
               #1% - insert title text
               \par% - end a paragraph (skip \parsep)
               \egroup% - terminate local group
              }% - complete definition

\renewcommand{\section}% - name of command
              [1]% - number of arguments
              {% - begin definition
               \bgroup% - begin a local group
               \flushleft% - turn on flushleft option
               \small\bf% - select font
               \refstepcounter{section}% - increment counter
               \arabic{section}. #1% - insert title text
               \par% - end a paragraph (skip \parsep)
               \egroup% - terminate local group
              }% - complete definition

\renewcommand{\subsection}% - name of command
              [1]% - number of arguments
              {% - begin definition
               \bgroup% - begin a local group
               \flushleft% - turn on flushleft option
               \small\em% - select font
               \refstepcounter{subsection}% - increment counter
               \arabic{section}.% - insert title text
               \arabic{subsection}. #1% - insert title text
               \par% - end a paragraph (skip \parsep)
               \egroup% - terminate local group
              }% - complete definition

\renewcommand{\subsubsection}% - name of command
              [1]% - number of arguments
              {% - begin definition
               \bgroup% - begin a local group
               \flushleft% - turn on flushleft option
               \small\em% - select font
               \refstepcounter{subsubsection}% - increment counter
               \arabic{section}.% - insert title text
               \arabic{subsection}.% - insert title text
               \arabic{subsubsection}. #1% - insert title text
               \par% - end a paragraph (skip \parsep)
               \egroup% - terminate local group
              }% - complete definition

  \newcommand{\acknowledgement}% - name of command
              [1]% - number of arguments
              {% - begin definition
               \bgroup% - begin a local group
               \flushleft% - turn on flushleft option
               \small\bf% - select font
               #1% - insert title text
               \par% - end a paragraph (skip \parsep)
               \egroup% - terminate local group
              }% - complete definition

  \newcommand{\sectionbib}% - name of command
              [1]% - number of arguments
              {% - begin definition
               \bgroup% - begin a local group
               \flushleft% - turn on flushleft option
               \small\bf% - select font
               #1% - insert title text
               \par% - end a paragraph (skip \parsep)
               \egroup% - terminate local group
              }% - complete definition

\begin{document}

% -------------------------------------------------------------------- %
% -------------------------------------------------------------------- %
% -------------------------------------------------------------------- %

% -------------------------------------------------------------------- %

\small
\baselineskip 10pt

% -------------------------------------------------------------------- %
% -------------------------------------------------------------------- %
% -------------------------------------------------------------------- %
\setcounter{page}{1}
% -------------------------------------------------------------------- %
\title{\LARGE \bf Modeling subgrid scale production rates on \\
complex meshes using graph neural networks}

% \author{{\large Author 1 full name$^{a,*}$, Author 2 full name$^{a,b}$, Author 3 full name$^{b}$, $\ldots$}\\[10pt]
%         {\footnotesize \em $^a$Author affiliation 1}\\[-5pt]
%         {\footnotesize \em $^b$Author affiliation 2}\\[-5pt]
%         {\footnotesize \em Continue the list of affiliations as needed, with one per line}}

\author{{\large Priyabrat Dash$^{a,*}$, Mathis Bode$^{b}$, Konduri Aditya$^{a}$}\\[10pt]
        {\footnotesize \em $^a$Department of Computational and Data Sciences, Indian Institute of Science, Bengaluru, 560012, KA, India}\\[-5pt]
        {\footnotesize \em $^b$Jülich Supercomputing Centre, Forschungszentrum Jülich GmbH, Jülich, 52425, NRW, Germany}\\[-5pt]
        }

\date{}  %%% Leave as is, do not add date;

% -------------------------------------------------------------------- %
% -------------------------------------------------------------------- %
% -------------------------------------------------------------------- %
\twocolumn[\begin{@twocolumnfalse}
\maketitle
\rule{\textwidth}{0.5pt}
\vspace{-5pt}

\begin{abstract} 
Large-eddy simulations (LES) require closures for filtered production rates because the resolved fields do not contain all correlations that govern chemical source terms. We develop a graph neural network (GNN) that predicts filtered species production rates on non-uniform meshes from inputs of filtered mass fractions and temperature. Direct numerical simulations of turbulent premixed hydrogen-methane jet flames with hydrogen fractions of 10\%, 50\%, and 80\% provide the dataset. All fields are Favre filtered with the filter width matched to the operating mesh, and learning is performed on subdomain graphs constructed from mesh-point connectivity. A compact set of reactants, intermediates, and products is used, and their filtered production rates form the targets. The model is trained on 10\% and 80\% blends and evaluated on the unseen 50\% blend to test cross-composition generalization. The GNN is compared against an unclosed reference that evaluates rates at the filtered state, and a convolutional neural network baseline that requires remeshing. Across in-distribution and out-of-distribution cases, the GNN yields lower errors and closer statistical agreement with the reference data. Furthermore, the model demonstrates robust generalization across varying filter widths without retraining, maintaining bounded errors at coarser spatial resolutions. A backward facing step configuration further confirms prediction efficacy on a practically relevant geometry. These results highlight the capability of GNNs as robust data-driven closure models for LES on complex meshes.
\end{abstract}

\vspace{10pt}

{\bf Novelty and significance statement}

\vspace{10pt}

This study introduces a mesh-native graph neural network (GNN) closure for predicting filtered species production rates directly on non-uniform meshes. By operating on discrete mesh connectivity, this framework avoids interpolation and remeshing steps required by conventional convolutional neural networks (CNNs). The model learns the mapping between filtered species mass fractions, temperature, and production rates using direct numerical simulation data from premixed \ce{H2}/\ce{CH4} jet flames filtered to LES resolution. Across both in-distribution and out-of-distribution \textit{a priori} evaluations, including an unseen intermediate fuel blend, the GNN achieves lower prediction errors and closer joint probability agreement with the reference compared to a matched-complexity CNN and a no-model baseline. This approach further demonstrates robustness to spatial resolution variations, maintaining accurate predictions across coarser filter widths without retraining. By enabling subgrid chemistry closure directly on solver meshes, this interpolation-free framework provides a scalable pathway for data-driven finite-rate chemistry modeling in LES of practical combustors.

\vspace{5pt}
\parbox{1.0\textwidth}{\footnotesize {\em Keywords:} turbulent reacting flows; data-driven combustion modeling; subgrid chemistry; graph neural networks}
\rule{\textwidth}{0.5pt}
*Corresponding author.
\vspace{5pt}
\end{@twocolumnfalse}] 

% \linenumbers
\section{Introduction \label{sec:introduction}} \addvspace{10pt}

Large-eddy simulations (LES) are widely used in studies of turbulent combustion because they offer a practical balance between accuracy and computational cost. Such calculations resolve the energy-containing large-scale motions and relevant unsteady structures while modeling the small scales, enabling simulations of realistic configurations that are prohibitive with direct numerical simulations (DNS). In LES, the Favre-filtered species transport equations are advanced in time, and accurate representation of the chemical source term is difficult because reactions occur in thin fronts that are typically smaller than the mesh size. The filtered equations therefore contain source terms that depend on correlations between unresolved thermochemical fluctuations and the resolved state. At subgrid scales, turbulence and chemistry interact in a coupled manner that redistributes scalars, modifies the local temperature, and potentially alters reaction pathways, so the resolved fields alone do not determine the filtered reaction rates. Consequently, the filtered production rate is not equal to the reaction rate evaluated at the filtered thermochemical state, i.e., $\widetilde{\dot{\omega}}_k \neq \dot{\omega}_k(\tilde{\phi})$. A closure is therefore required to quantify the effect of these subgrid {(or subfilter)} correlations, and the predictive capability of LES depends strongly on the quality of this closure. Since operating conditions, combustion regimes, and transported scalars vary across applications, no single closure performs consistently across configurations.

 Conventional closure models address this problem through chemistry tabulation or statistical descriptions of unresolved fluctuations. Flamelet-based approaches \cite{Peters1988}, including flamelet/progress-variable formulations \cite{Pierce2004} and flamelet-generated manifolds \cite{Oijen2000}, reconstruct reaction rates by parameterizing the local thermochemical state using a reduced set of control variables \citep{Wick2020,Gierth2022}. In premixed combustion, flame surface density (FSD) \cite{Trouve1994} and artificially thickened flame (TFLES) \cite{Colin2000} models are frequently employed to resolve subgrid-scale flame-turbulence interactions. These formulations typically rely on algebraic wrinkling factor closures \cite{Charlette2002} to account for turbulence-induced flame surface enhancement. Alternative formulations rely on statistical descriptions of subgrid fluctuations. Conditional moment closure \cite{Bilger1993} and transported probability density function methods \cite{Pope1991} model the joint statistics of thermochemical variables to compute filtered reaction rates. 

Other finite-rate chemistry closures, such as the partially stirred reactor (PaSR) \cite{Magnussen1981} and eddy dissipation concept (EDC) \cite{Pequin2022} models, represent subgrid mixing and reactions through local reactor analogies and modeled turbulence timescales. {Hybrid parameterizations} \cite{Ihme2010} combining mixture fraction, progress variables, and scalar dissipation rates improve tabulation accuracy across different combustion regimes. While these methods remain widely used in practical simulations, their predictive capability deteriorates when the assumed flame structure, reduced thermochemical parameterization, or modeled scalar statistics fail to represent local turbulence-chemistry interactions. Furthermore, classical closures estimate filtered reaction rates primarily from the local thermochemical state. They incorporate spatial correlations present in the resolved flow field only indirectly through modeled quantities such as the scalar dissipation rate or wrinkling factors.

Recent works have explored data-driven closures for turbulent combustion using neural networks trained on DNS fields filtered to LES resolutions. First-generation data-driven closures employed multi-layer perceptrons to approximate filtered reaction rates from local thermochemical scalars, providing a direct mapping between the resolved state and subgrid chemistry \cite{Shin2022, Piu2025}. To embed established physics into these data-driven predictions, parallel efforts have developed hybrid frameworks that augment classical closures with machine learning, such as using neural networks to dynamically model subgrid variances for presumed PDF methods \cite{HenryDeFrahan2019}. While these pointwise and hybrid models demonstrate encouraging \emph{a priori} performance, they process each computational node independently, thereby incorporating spatial correlations in the resolved field only indirectly. To address this limitation, convolutional neural networks (CNNs) have been introduced to exploit spatial structure within filtered DNS data. Convolution-based architectures have shown strong performance in a range of reacting and non-reacting flow applications, reporting high \emph{a priori} accuracy for predicting filtered reaction rates and subgrid flame surface wrinkling \citep{Lapeyre2019,Male2025,Arumapperuma2025}. However, standard CNNs are formulated exclusively for structured uniform meshes where convolution operators assume regular grid connectivity. Practical combustor simulations on non-uniform or unstructured meshes therefore require field interpolation or remeshing, which can distort the thermochemical structure and introduce additional errors in predicted filtered production rates of species.

Graph neural networks (GNNs) overcome this topological limitation by learning operators directly on complex computational meshes \cite{Battaglia2018}. By performing localized message passing along the edges of a graph that maps the exact discrete connectivity, GNNs inherently account for varying mesh spacing and irregular topology without requiring flow-field interpolation. This mesh-native capability has driven the recent adoption of GNNs across fluid mechanics, including data-driven prediction of future flow states \cite{Lino2022}, interpretable compression of high-dimensional fields \cite{Barwey2023}, and super-resolution of turbulent non-reacting \cite{Barwey2025} and reacting \cite{Dash2026} flows. In the context of turbulence modeling, GNN-based architectures have been explored for subgrid-scale modeling in non-reacting flows, demonstrating improved robustness when generalizing beyond canonical configurations \cite{Kim2025}.

Many existing machine-learning closures \citep{Lapeyre2019,Male2025,Arumapperuma2025} for reacting flows focus on predicting a single scalar quantity, such as a progress-variable source term or flame surface density. In such formulations, the coupling {among} multiple thermochemical scalars is embedded implicitly within the chosen parameterization. However, practical combustion dynamics are governed by the simultaneous, highly nonlinear evolution of multiple species mass fractions and temperature. Directly predicting an extensive set of filtered production rates from the resolved multi-species state is therefore essential to capture these intricate thermochemical interactions accurately in LES. Furthermore, alternative reconstruct-then-evaluate strategies, which infer subgrid chemistry by first super-resolving the scalar fields and subsequently computing reaction rates, incur substantial computational overheads.  While these approaches can provide valuable physical insight \cite{Nista2025}, the combined computational overhead of performing the super-resolution inference itself and subsequently evaluating detailed chemical kinetics on the reconstructed high-resolution grid makes them significantly expensive. This hinders their efficient integration into simulations of practical combustors.

This study takes a direct, mesh-native, data-driven approach toward modeling subgrid chemistry. We learn the mapping between the filtered thermochemical state and the filtered species production rates on the native structured non-uniform mesh to demonstrate \emph{a priori} subgrid-scale (SGS) modeling. The graph neural network predicts filtered production rates from filtered species mass fractions and temperature using localized message passing that preserves mesh connectivity. Although the present work focuses on structured non-uniform meshes, the graph formulation does not assume a particular mesh topology and can in principle extend to unstructured meshes. This setup isolates the closure model's behavior from errors introduced by a coupled LES solver, such as spatial and temporal discretization or auxiliary subgrid closures.  Finally, to quantify its predictive capability and generalization, we systematically compare the GNN predictions against {an unclosed no-model baseline (resolved-state evaluation)} and a data-driven CNN baseline.

\section{Methodology\label{sec:methodology}} \addvspace{10pt}

\begin{figure*}[h!]
\centering
\setlength{\unitlength}{1cm}

\includegraphics[trim={0cm 0cm 0cm 0cm},clip,width=12cm]{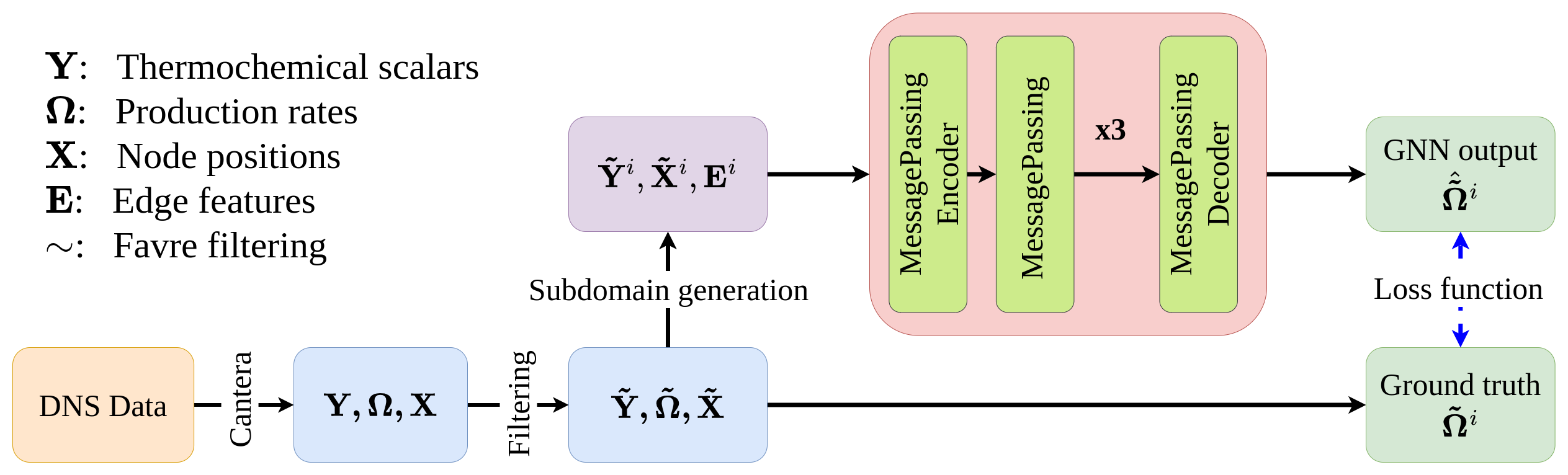}

\caption{\footnotesize A schematic of the GNN framework for modeling subgrid-scale production rates.}
\label{fig:schematic}
\end{figure*}

The mass conservation equation of a reacting-species $\alpha$ expressed in terms of mass fraction \(Y_\alpha\) is
\[
\frac{\partial \rho Y_\alpha}{\partial t}
+\nabla\cdot(\rho \mathbf{u} Y_\alpha)
= -\,\nabla\cdot(\rho \mathbf{V}_\alpha Y_\alpha)
+ \dot{\omega}_\alpha,
\]
where \(\rho\) is the density, \(\mathbf{u}\) is the velocity, and \(\mathbf{V}_\alpha\) and \(\dot{\omega}_\alpha\) are the diffusive velocity and the production rate of species \(\alpha\), respectively. After Favre filtering (\(\tilde{\phi}=\overline{\rho \phi}/\bar{\rho}\)), the filtered transport equation contains the filtered source \(\widetilde{\dot{\omega}}_\alpha\) and additional unclosed correlations in the convective and diffusive fluxes. The {highly} nonlinear dependence of reaction rates on temperature ($\tilde{T}$), pressure ($\bar{p}$), and species mass fractions ($\tilde{\mathbf{Y}}$) dictates that the filtered source term cannot be recovered by evaluating the chemical mechanism with filtered scalars, meaning $\widetilde{\dot{\omega}}_\alpha \neq \dot{\omega}_\alpha(\tilde{\mathbf{Y}},\tilde{T},\bar{p})$. This mismatch defines the subgrid chemistry closure problem, and motivates the development of models that can recover filtered production rates from the resolved thermochemical state. To address this, the present work develops a data-driven model to map the resolved thermochemical state to the filtered production rates. Particularly, this mapping is learned directly on the native structured non-uniform mesh in this study and can obviously be extended to unstructured meshes. Avoiding interpolation of data onto a uniform Cartesian grid ensures that the closure remains consistent with the spatial discretization used {in} practically-relevant large eddy simulations (LES).

A schematic of the proposed methodology is shown in Fig.~\ref{fig:schematic}. It starts with high-resolution DNS fields of density, temperature, species mass fractions, and production rates obtained with a detailed chemical mechanism. From these fields, the filtered thermochemical states and production rates are computed by applying Favre filtering, so that the resulting dataset emulates LES fields at the target resolution. The filter width in this study is set to be equal to the local spacing of the target LES mesh. Additionally, the entire methodology operates on stored DNS time instances and a thermochemistry library such as \texttt{Cantera} for evaluating production rates. Hence, it does not require any modification of, or direct access to an LES solver in this \textit{a priori} setting for training and inference. Nevertheless, this methodology naturally extends to \textit{a posteriori} deployment. A trained model can serve as a drop-in subgrid closure as the native state variables of an LES solver inherently represent the required Favre-filtered quantities. To accommodate single-GPU (Nvidia A100, for reference) memory limits, the three-dimensional fields are spatially partitioned into non-overlapping subdomains. These subdomains are the basic units for graph construction, training, and inference.

%Overall graph structure in words
For computing closures directly on the native mesh, we reformulate each subdomain as a graph. Mesh points are treated as nodes, with node positions given by the corresponding cell centers. A node neighborhood is established using one-hop connections along the mesh coordinate directions {in three dimensions}, yielding a standard six-point stencil plus a self-connection. Consequently, interior nodes possess exactly seven edges, while boundary nodes possess fewer edges based on their location within the subdomain. For subdomain $i$, the graph is denoted as $\mathcal{G}^{\,i} = (V^{\,i}, \mathbf{E}^{\,i})$, where $V^{\,i}$ is the node set and $\mathbf{E}^{\,i}$ is the edge set.

%Mathematical description
For each mesh point (node), the model inputs consist of a selected subset of filtered species mass fractions together with the filtered temperature. Unlike many prior data-driven subgrid-scale closures that rely on a single progress variable or a reduced manifold (e.g., progress variable and mixture fraction), the present approach employs multiple species to represent the local thermochemical state. This choice limits computational cost while retaining the principal reactants, intermediates, and products necessary to map the local thermochemical state to the filtered reaction rates. For a given subdomain \(i\), the filtered thermochemical state is represented by a node-feature matrix 
\(\tilde{\mathbf{Y}}^{\,i} \in \mathbb{R}^{N_n \times N_s}\), where \(N_n = 32^3\) denotes the number of local mesh points and \(N_s\) corresponds to the number of input scalars, comprising filtered species mass fractions and temperature. The physical coordinates of the nodes are stored in \(\tilde{\mathbf{X}}^{\,i} \in \mathbb{R}^{N_n \times 3}\), explicitly encoding the geometry of the native non-uniform mesh. The training targets consist of the filtered production rates of the same subset of species, represented for the corresponding subdomain by 
\(\tilde{\boldsymbol{\Omega}}^{\,i} \in \mathbb{R}^{N_n \times (N_s-1)}\). The corresponding model predictions are denoted by \(\hat{\tilde{\boldsymbol{\Omega}}}^{\,i}\).
The edge features between neighboring nodes $j$ and $k$ are initialized in a form consistent with the \texttt{PyTorch Geometric} framework as
$\boldsymbol{\xi}^{\,i}_{jk}
= \big[\tilde{\mathbf{Y}}^{\,i}_{j} - \tilde{\mathbf{Y}}^{\,i}_{k},\ 
   \tilde{\mathbf{X}}^{\,i}_{j} - \tilde{\mathbf{X}}^{\,i}_{k},\ 
   \|\tilde{\mathbf{X}}^{\,i}_{j} - \tilde{\mathbf{X}}^{\,i}_{k}\|_2\big]$.
These edge feature vectors provide the network with local spatial gradients by encoding the differences in filtered scalars, relative coordinate offsets, and the Euclidean distance between connected nodes. The output graph for the same subdomain uses the same nodes and edges, but replaces the node features with the filtered production rates $\hat{\tilde{\boldsymbol{\Omega}}}^{\,i}$.

We employ an encoder-processor-decoder architecture with five message-passing (MP) layers, following the formulation of Pfaff et al.~\citep{Pfaff2021}. In each MP layer, information is exchanged between neighboring nodes through the graph connectivity defined previously. The filtered production rates depend not only on the local thermochemical state at each mesh point but also on the surrounding spatial context. This spatial coupling is particularly important in turbulent premixed flames, where reaction zones thinner than the LES filter width couple the filtered source term at a point with the subgrid scalar distribution in the surrounding neighborhood. MP layers enable the propagation of this spatial information across mesh points while accounting for the native non-uniform mesh spacing through localized edge and node feature updates.

For subdomain $i$ and layer $l$, we first update the edge features from the previous layer. The incoming edge features $\boldsymbol{\xi}_{jk}^{\,i,l-1}$, source node features $\boldsymbol{\phi}_{j}^{\,i,l-1}$, and target node features $\boldsymbol{\phi}_{k}^{\,i,l-1}$ are concatenated and passed through an edge multi-layer perceptron (MLP), denoted $\mathrm{MLP}^{\,l}_{e}$: $
\boldsymbol{\xi}_{jk}^{\,i,l}
=
\mathrm{MLP}^{\,l}_{e}
\left(
\boldsymbol{\xi}_{jk}^{\,i,l-1},
\boldsymbol{\phi}_{j}^{\,i,l-1},
\boldsymbol{\phi}_{k}^{\,i,l-1}
\right)$.
Next, the updated edge features are aggregated over the neighborhood $\mathcal{N}(j)$ of each node $j$ using a permutation-invariant mean aggregation:
$\mathbf{h}_{j}^{\,i,l} = \frac{1}{|\mathcal{N}(j)|} \sum_{k \in \mathcal{N}(j)} \boldsymbol{\xi}_{jk}^{\,i,l}$.
Mean aggregation is chosen over sum aggregation to ensure independence on the number of neighbors. The aggregated messages $\mathbf{h}_{j}^{\,i,l}$ are concatenated with the previous node features and passed through a node MLP, $\mathrm{MLP}^{\,l}_{\phi}$:
$\boldsymbol{\phi}_{j}^{\,i,l} = \mathrm{MLP}^{\,l}_{\phi}\left(\mathbf{h}_{j}^{\,i,l}, \boldsymbol{\phi}_{j}^{\,i,l-1}\right)$.
This sequence of edge update, message aggregation, and node update is repeated across five layers. With the six-point stencil connectivity (two along each direction), each layer communicates information over one additional hop. Thus, five layers enable interior nodes to access information from up to a five-hop neighborhood, which corresponds to a physical distance of approximately five times the local mesh spacing. This receptive field is comparable to the integral length scale at LES resolution and sufficient to capture the dominant spatial correlations in filtered {production} rates, while remaining computationally tractable for $32^3$ subdomains.

Prior to training, we scale all input and target variables using min-max normalization computed separately for each scalar from the training set statistics. This prevents systematic over- or under-prediction of species with disparate production-rate magnitudes, which is common in combustion due to the wide dynamic range across reactants, intermediates, and products. Each scalar is mapped to the range $[0,1]$, thus ensuring balanced gradient contributions during optimization.

All MLPs are configured with {two hidden layers}, each comprising of 48 neurons, exponential linear unit (ELU) activation for intermediate layers, and layer normalization after each node update. The hidden layer width of 48 is selected based on individual GPU memory. The two-hidden-layer structure in each MP layer provides sufficient nonlinearity for the thermochemical mappings without excessive parameter count. ELU activation provides smooth behavior for negative inputs and avoids the dead-neuron problem that can stall training, an advantage previously demonstrated to enhance gradient propagation in deep message-passing architectures. To strictly bound the network predictions to the $[0,1]$ normalized ground truth space, a sigmoid activation function is applied at the final output layer. Layer normalization is applied after each node feature update to address two combustion-specific challenges. First, message magnitudes vary significantly between flame fronts (high gradients), and {pre- and post-flame regions} (near-equilibrium). Second, boundary nodes have fewer neighbors than interior nodes, leading to inconsistent aggregation scales. Layer normalization stabilizes training across these heterogeneities.

The forward propagation process of the model (with parameters $\theta$) is expressed as $\hat{\tilde{\boldsymbol{\Omega}}}^{\,i} = \mathrm{GNN}(\tilde{\mathbf{Y}}^{\,i}, \tilde{\mathbf{X}}^{\,i}, \mathbf{E}^{\,i}; \theta)$.
The model is trained by minimizing the mean-squared error (MSE) between predicted and reference filtered production rates. For subdomain $i$, the loss is computed as
$
\mathcal{L}^i
=
\frac{1}{N_n}
\left\|
\hat{\tilde{\boldsymbol{\Omega}}}^{\,i}
-
\tilde{\boldsymbol{\Omega}}^{\,i}
\right\|_2^2$,
where $N_n$ is the number of nodes per subdomain. {The model is implemented in \texttt{PyTorch Geometric} \cite{Fey2025}, and its parameters are updated using the Adam optimizer on a multi-node multi-GPU platform with 16 NVIDIA A100 GPUs}. The subdomains generated after Favre filtering are randomly split into training (90\%) and validation (10\%) sets. The learning rate is manually reduced from an initial value of $1\times10^{-4}$ upon stagnation of the validation loss for 100 epochs.

For establishing a data-driven baseline, we train a convolutional neural network (CNN) utilizing the identical filtered dataset. The CNN is matched to the GNN in number of layers (five convolutional blocks) and intermediate feature width (48 channels), so performance differences reflect only the architectural choice rather than model capacity. Notably, standard convolution layers assume uniform spatial discretization. Consequently, the CNN workflow requires interpolating the non-uniform mesh data onto a uniform $32^3$ mesh for training and inference, followed by an inverse interpolation of the predictions back to the native mesh. This baseline comparison directly isolates the specific advantage of the GNN in preserving exact native-mesh connectivity without interpolation errors.

\section{Datasets\label{sec:datasets}} \addvspace{10pt}
\subsection{Turbulent premixed \ce{H2}-\ce{CH4} jet flames} \addvspace{10pt}
We first develop and evaluate the model using three-dimensional DNS of unconfined, turbulent premixed hydrogen-methane jet flames computed with \texttt{NTMIX-CHEMKIN} \cite{Ho2024}. The dataset \cite{Chung2022} comprises {stoichiometric} blends containing 10\%, 50\%, and 80\% hydrogen by volume (H10, H50, H80). The bulk flow conditions are matched across all cases by maintaining a constant jet Reynolds number of 10,300 and {turbulent inflow}. To resolve the multiscale flame dynamics while damping acoustic reflections, the spatial discretization incorporates explicit grid stretching towards the open boundaries, yielding natively non-uniform meshes comprising up to one billion mesh points. Chemistry is modeled using a 16-species reduced mechanism derived from the Foundational Fuel Chemistry Model (FFCM-1.0). The selected species subset for SGS modeling focuses on fuels ($\ce{H2}$, $\ce{CH4}$), key intermediates ($\ce{OH}$, $\ce{HO2}$, $\ce{CH2O}$, $\ce{CO}$), and products ($\ce{CO2}$, $\ce{H2O}$).

The dataset is constructed by selecting four discrete time instants from the H10 and H80 cases for training, while the remaining instants are reserved for in-distribution testing. As the hydrogen concentration increases, the Karlovitz number (\(Ka\), defined as the ratio of the flame time scale to the Kolmogorov time scale) decreases from 70.4 in H10 to 19.2 in H80. The model is trained for 5150 epochs. To assess out-of-distribution generalization, the trained model is further evaluated on an unseen time instant from the intermediate H50 case (\(Ka = 38.2\)). This test probes the model’s ability to generalize across variations in fuel composition and flame stretch, while remaining agnostic to the intermediate blend during training.

\begin{figure*}[t]
\centering
\setlength{\unitlength}{1cm}
\includegraphics[trim={0cm 0cm 0cm 0cm},clip,width=14cm]{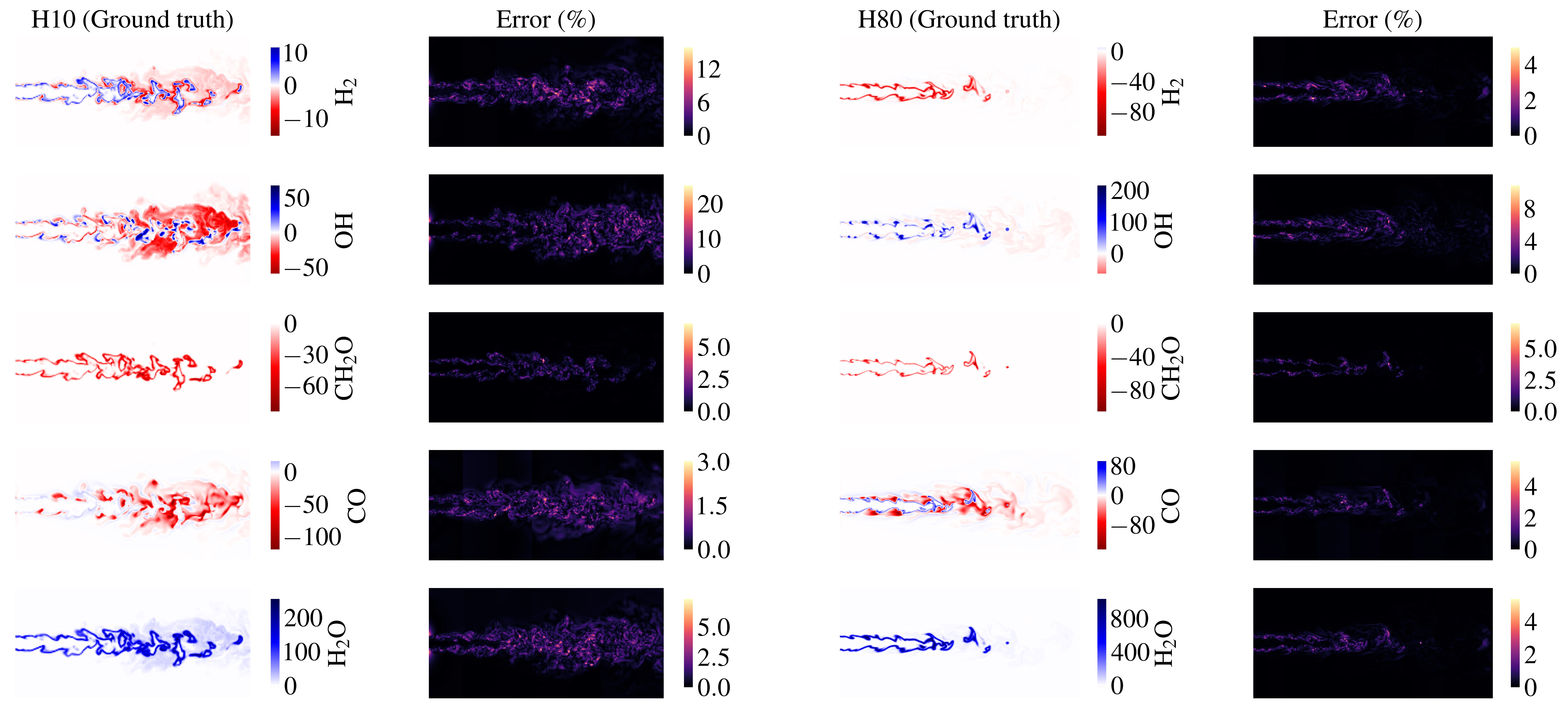}

\caption{\footnotesize Contours of GNN predictions for in-distribution test cases. \textbf{Columns 1 and 2} display the H10 case, showing species production rates (in $\mathrm{kg \ m^{-3} \ s^{-1}}$) at the domain midplane and corresponding percentage errors relative to filtered DNS data, respectively. \textbf{Columns 3 and 4} present the same quantities for the H80 test case.}
\label{fig:in-dist-err-contours}
\end{figure*}

\subsection{Lean-premixed \ce{C2H4}-air flame stabilized over a backward facing step} \addvspace{10pt}
For demonstrating versatility, we evaluate the closure model on a dataset corresponding to a turbulent reacting flow past a backward facing step (BFS) combustor \cite{Aditya2019, Dash2023}. The BFS geometry serves as a fundamental building block for cavity-based flameholders in scramjets and gas turbines, where sudden flow expansion generates the necessary aerodynamic recirculation zones. The high-resolution dataset, generated using the multi-block \texttt{S3D} solver, features a lean premixed ethylene-air (\ce{C2H4}-air) flame. 
Here, we focus on a spatially non-uniform subdomain containing the shear layer and recirculation zone. Within this region, the original DNS mesh utilizes a uniform $12.5\,\mu\mathrm{m}$ mesh size streamwise and spanwise, but is stretched transversely (from $8\,\mu\mathrm{m}$ near the wall to $20\,\mu\mathrm{m}$ along the combustor centerline). It is emphasized that this assessment is not an exercise in zero-shot geometric extrapolation. Rather, the identical GNN architecture is independently trained using the BFS dataset {to assess its capability to adapt to} different combustor geometries, stabilization mechanisms, and chemical kinetics while preserving the native non-uniform mesh structure. {The selected species subset comprises reactants ($\ce{C2H4}$, $\ce{O2}$), key intermediates ($\ce{OH}$, $\ce{HO2}$, $\ce{CH2O}$, $\ce{CO}$), and products ($\ce{CO2}$, $\ce{H2O}$). The GNN is trained for 3000 epochs in this case.}

\section{Results\label{sec:results}} \addvspace{10pt}

The trained graph neural network is evaluated on unseen time instances of unconfined turbulent jet flames to assess its generalization across thermochemical states. The model is trained using four instances each from the H10 and H80 datasets, where locally, the mesh size is 8$\times$ that of the DNS mesh while maintaining non-uniformity. During full-domain inference, subdomains are augmented with a one-node-thick buffer region to maintain complete neighborhood stencils, ensuring consistent message-passing operations and mitigating windowing artifacts. Figure~\ref{fig:in-dist-err-contours} presents contours of the filtered production rates (ground truth) and the corresponding errors incurred by the GNN (computed as a percentage of peak absolute production rate). Absolute production rates in the H80 flame are notably higher due to increased mixture reactivity. Despite these significant disparities in reaction intensity, the model limits relative errors to within 10\% across the majority of the domain for both flames. The largest discrepancies are restricted to the highly reactive \ce{OH} radical, which exhibits localized erroneous regions bound by 20\%. The spatial distribution of these errors correlates directly with the respective flame structures. The methane-dominated H10 flame produces a longer flame brush, leading to spatially extended production zones for intermediates. The GNN effectively identifies these elongated pathways, keeping \ce{CH2O} and \ce{CO} errors below 5\% and successfully capturing distributed subgrid correlations driven by the slower carbon oxidation chemistry. Conversely, the high hydrogen content in H80 accelerates kinetics, confining heat release to thin reaction fronts. The GNN accurately captures these intense gradients, restricting \ce{H2} and \ce{OH} errors to narrow pockets. Finally, errors for \ce{H2O} (a product species) remain below 5\% in both cases, indicating that the approach reliably recovers the overall product formation rates for both mixtures.

\begin{figure*}
\centering
\setlength{\unitlength}{1cm}
\includegraphics[trim={0cm 0cm 0cm 0cm},clip,width=14cm]{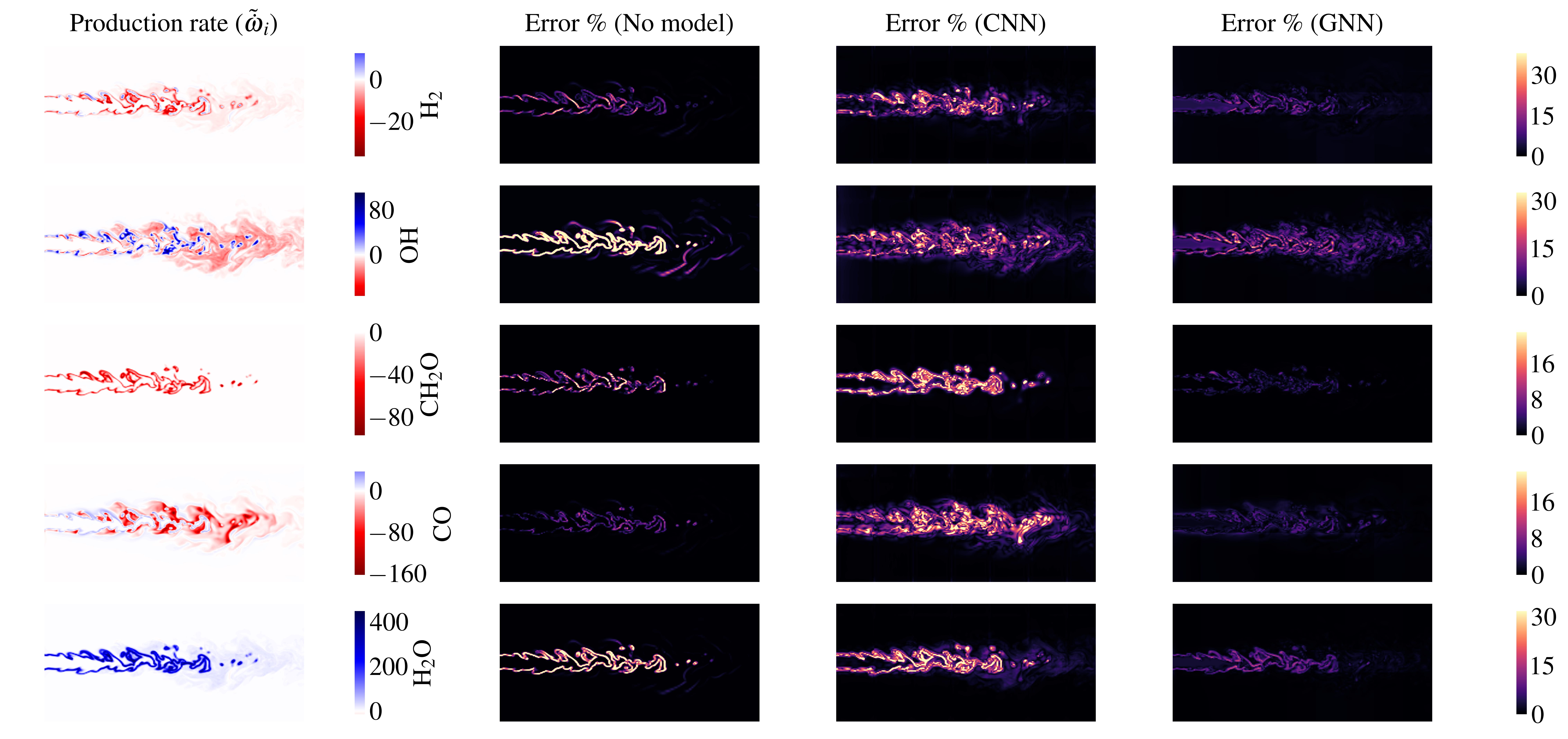}
% \caption{\footnotesize }
\caption{\footnotesize Contours demonstrating superiority of GNN compared to other approaches {for H50}. {\bf Column 1:} Contours of production rates of key species (in \(\mathrm{kg\,m^{-3}\,s^{-1}}\)) shown for the domain midplane. {\bf Columns 2, 3, 4:} Contours of percentage errors between ground truth (filtered from DNS solution) and the results of no-model, CNN, and GNN.}
\label{fig:ood-err-contours}
\end{figure*}

Evaluating out-of-distribution performance is essential for practical subgrid-scale modeling, as operating conditions in combustors continuously vary. To assess generalization across \ce{H2}/\ce{CH4} fuel blends, the trained model is evaluated on the previously unseen H50 flame dataset. Due to the nonlinear nature of chemical kinetics, the intermediate 50\% \ce{H2} blend does not manifest as a simple linear superposition of the training data. {Furthermore, it operates at a distinct Karlovitz number (\(Ka = 38.2\)) compared to the H10 (\(Ka = 70.4\)) and H80 (\(Ka = 19.2\)) training conditions.} Additionally, this H50 case is also discretized with a different non-uniform mesh. To quantify accuracy, the GNN is compared against two baselines: an unclosed no-model reference that evaluates detailed chemistry directly at the filtered state, and an equivalently complex CNN that requires trilinear interpolation to a uniform Cartesian mesh for inference. {Comparisons against conventional algebraic closures are omitted, as recent studies \cite{Male2025, Arumapperuma2025} have established that CNN-based architectures consistently perform better than classical subgrid models.}

Figure~\ref{fig:ood-err-contours} compares the spatial error distributions of the three approaches. The no-model evaluation exhibits large discrepancies exceeding 100\%, localized predominantly within the thin reaction zones. For clarity, the contour colorbars are intentionally capped at values well below the peak errors of the no-model case to better highlight the differences between the closures. Evaluating production rates using only the resolved fields fundamentally ignores the turbulence-chemistry interactions occurring at the subgrid level, confirming that explicit closure is mandatory. The CNN baseline attenuates these peak errors but distributes them across a much broader spatial extent. This artificial smearing stems directly from the required interpolation away from the native mesh, thereby sacrificing the critical non-uniform refinement clustered near the flame front. The GNN circumvents these topological constraints, maintaining prediction errors below 10\% for most species. The largest errors are again observed for \ce{OH}. The production rate of \ce{OH} is highly localized, features rapid sign changes over short distances, and is intensely temperature-dependent. These fast radical kinetics exponentially amplify minute perturbations in the filtered input states, which physically accounts for the higher prediction errors observed for this species. More broadly, the GNN strictly limits its predictions across all species to the chemically active jet and predicts precisely zero production in the unburned far-field. Preventing spurious source terms in chemically inactive regions is an essential requirement for maintaining global stability and preventing artificial ignition in \textit{a posteriori} applications.

\definecolor{gnnblue}{RGB}{100, 145, 190} 
\definecolor{cnnred}{RGB}{210, 100, 90}

\begin{figure}[h!]
  \centering
  \setlength{\unitlength}{1cm}
  \hspace{-0.5cm}
  \begin{picture}(7,5.5)
    % Base image
    \put(0,0){\includegraphics[trim={0cm 0cm 0cm 0cm},clip,width=7cm]{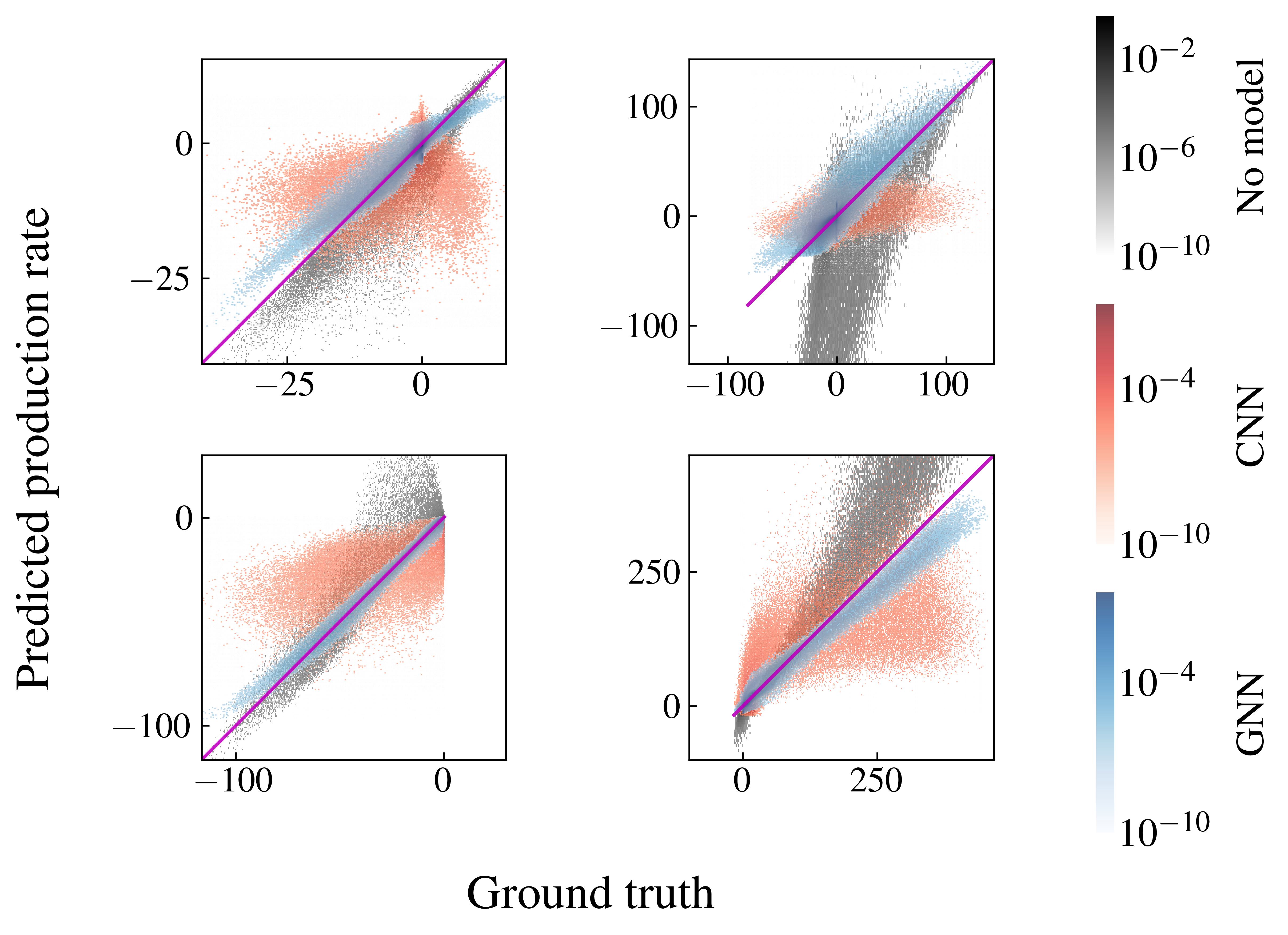}}
    
    % --- Top-Left: H2 ---
    \put(1.3, 4.4){\scriptsize \ce{H2}}
    \put(1.0, 4.8){\scriptsize [0.817 \textcolor{cnnred}{0.254} \textcolor{gnnblue}{0.896}]}
    
    % --- Top-Right: OH ---
    \put(3.9, 4.4){\scriptsize \ce{OH}}
    \put(3.65, 4.8){\scriptsize [-5.53 \textcolor{cnnred}{0.290} \textcolor{gnnblue}{0.787}]}
    
    % --- Bottom-Left: CH2O ---
    \put(1.3, 2.25){\scriptsize \ce{CH2O}}
    \put(1.0, 2.65){\scriptsize [0.926 \textcolor{cnnred}{0.647} \textcolor{gnnblue}{0.994}]}
    
    % --- Bottom-Right: H2O ---
    \put(3.9, 2.25){\scriptsize \ce{H2O}}
    \put(3.65, 2.65){\scriptsize [0.509 \textcolor{cnnred}{0.712} \textcolor{gnnblue}{0.974}]}
  \end{picture}
  \caption{\footnotesize Joint PDFs computed between production rates obtained from no-model/CNN/GNN and ground truth. Magenta line represents the $y=x$ reference curve, within the extents of GNN prediction. $R^2$ score mentioned in subplot title (no-model/CNN/GNN).}
  \label{fig:ood-jointprob}
\end{figure}
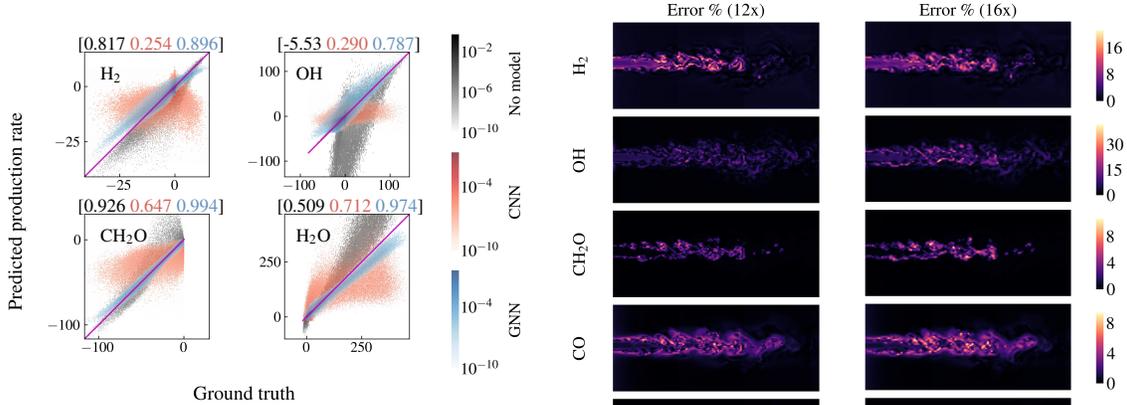

Beyond error contours, Fig.~\ref{fig:ood-jointprob} compares the joint probability density functions of the predicted and filtered production rates. The no-model approach exhibits severe distributional shifts, resulting in significant over-predictions of both \ce{OH} consumption and \ce{H2O} production. The CNN predictions fall within the correct global bounds but remain statistically diffuse and poorly aligned with the reference data. In contrast, the GNN tightly tracks the reference distribution across all scales. {This visual alignment is quantitatively confirmed by the computed $R^2$ scores. The proposed framework achieves scores exceeding $0.89$ for all species except the highly reactive \ce{OH} radical. The unclosed baseline entirely diverges with an $R^2$ of $-5.53$ and the CNN yields a weak correlation of $0.290$. The GNN effectively overcomes these limitations and maintains strong statistical fidelity with an $R^2$ of $0.787$.} The model successfully maps the nonlinear kinetics manifold, recovering the statistical structure of the turbulent flame directly on the non-uniform mesh despite the severe compositional shift.

\begin{figure}
% \centering
\setlength{\unitlength}{1cm}
\hspace{-0.5cm}
\includegraphics[trim={0cm 0cm 0cm 0cm},clip,width=7.5cm]{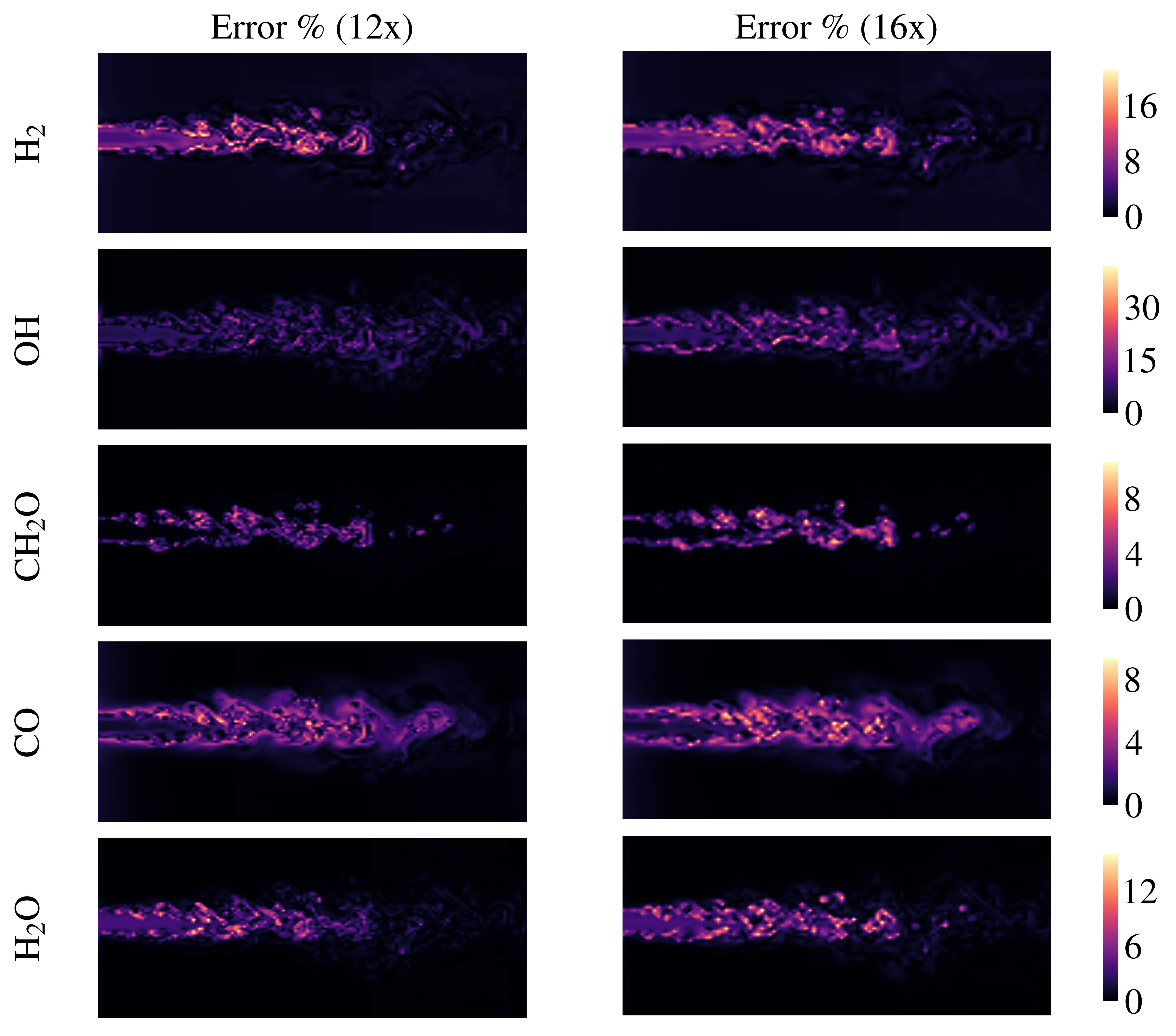}

\caption{\footnotesize Contours demonstrating the prediction error incurred by the GNN for the unseen H50 flame across varying filter widths. Percentage errors for the 12$\times$ downsampling factor are presented in {\bf Column 1} and for the 16$\times$ downsampling factor in {\bf Column 2}.}
\label{fig:ood-chem-downsamp}
\end{figure}

The next analysis evaluates model performance across spatial resolutions by predicting production rates for the unseen H50 dataset at downsampling factors of 12$\times$ and 16$\times$. Notably, the model was trained using data at a downsampling factor of 8$\times$. Such tests are relevant for LES, where filter sizes vary with computational constraints, and an effective closure should ideally retain predictive accuracy across mesh resolutions without retraining. Figure~\ref{fig:ood-chem-downsamp} shows the spatial error distributions for both cases. Barring the highly reactive \ce{OH} radical, prediction errors for all other species remain below 15\% across the domain at both resolutions. The GNN leverages local mesh connectivity to extract information from the strongly smoothed scalar fields and infer production rates despite simultaneous changes in fuel composition and filter size. As the downsampling factor increases from 12$\times$ to 16$\times$, both peak errors and the spatial extent of error regions increase. At such large filter widths, which significantly exceed the thermal flame thickness, most of the turbulent scalar {variation} and heat release occurs entirely within unresolved scales, leaving limited resolved gradients for the model. Despite this loss of resolved structure, the model maintains bounded prediction errors, indicating robustness for coarse-mesh LES where reaction zones are largely subgrid.

% \medskip
% \noindent\textbf{Backward facing step.}
\begin{figure}
\centering
\setlength{\unitlength}{1cm}
\hspace{-0.5cm}
\includegraphics[trim={0cm 0cm 0cm 0cm},clip,width=7.5cm]{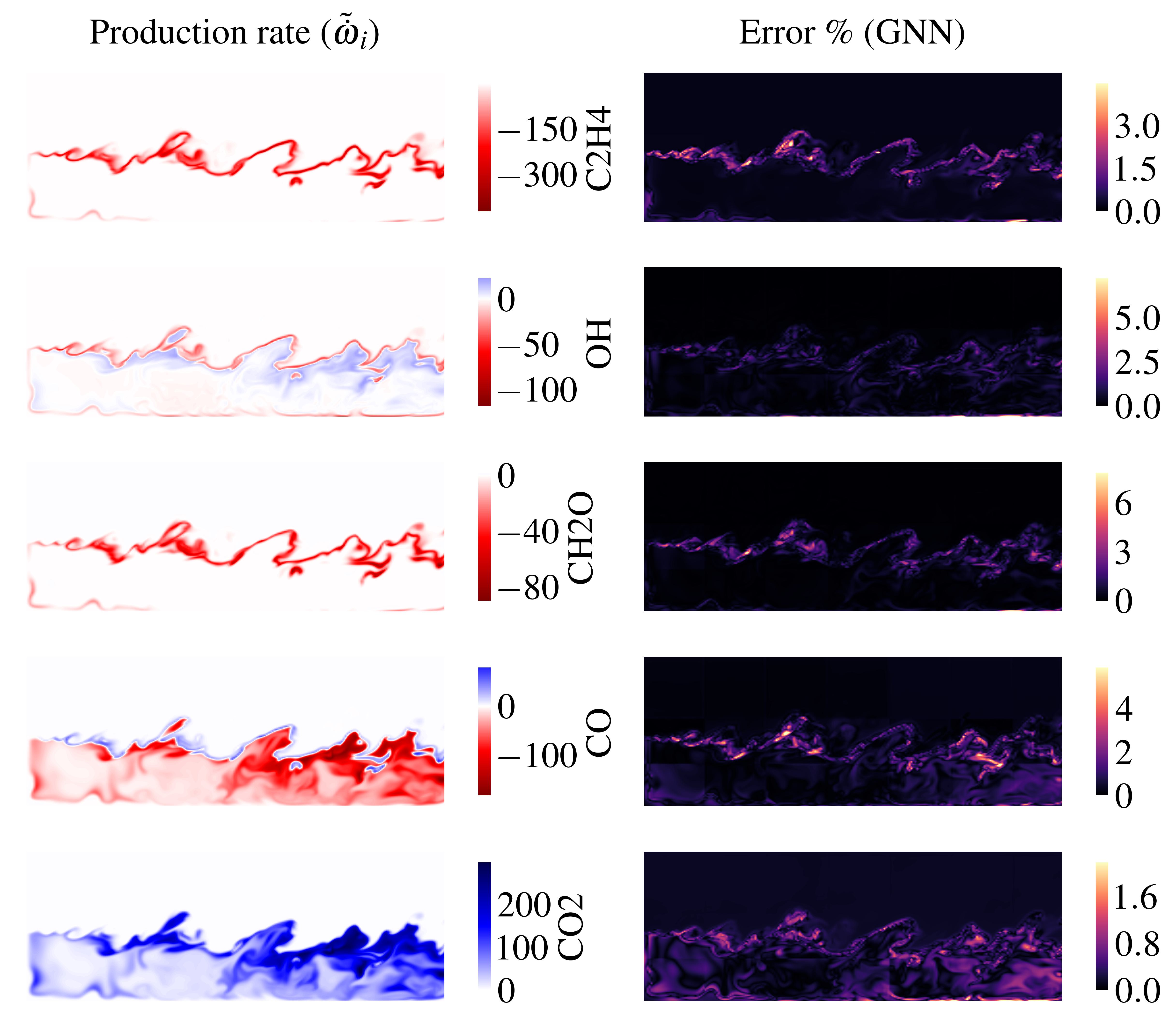}

\caption{\footnotesize Contours demonstrating the prediction error incurred by GNN. Ground truth (in \(\mathrm{kg\,m^{-3}\,s^{-1}}\)) is presented in {\bf Column 1} and percentage errors in {\bf Column 2}.}
\label{fig:bfs-err-contours}
\end{figure}

To evaluate the framework's applicability to configurations with increased geometric complexity, it is applied to a practically relevant case of a backward facing step featuring a turbulent lean premixed \ce{C2H4}-air flame \cite{Dash2023}. The flow field is characterized by flame stabilization within a highly wrinkled shear layer, introducing a recirculation zone and distinct turbulence-chemistry interactions not present in the unconfined jet flames. The analysis focuses on a subset of the combustor domain encompassing this shear region. Figure~\ref{fig:bfs-err-contours} presents the spatial error distributions for the fuel (\ce{C2H4}), key intermediates (\ce{OH}, \ce{CH2O}, \ce{CO}), and the primary product (\ce{CO2}). Across all species, the GNN demonstrates strong agreement with the filtered reference, successfully restricting discrepancies to below 10\% of the peak absolute production rate. The spatial contours reveal that prediction errors are tightly concentrated exclusively within the reacting shear layer, with near-zero spurious production elsewhere in the domain.

\section{Conclusions\label{sec:concl}} \addvspace{10pt}
In this study, we present a graph neural network that predicts filtered species production rates on non-uniform meshes from filtered mass fractions and temperature. Training uses hydrogen-methane jet flames at 10\% and 80\% hydrogen and testing uses an unseen 50\% blend. Across in-distribution and out-of-distribution cases, cumulative errors remain below 10\% for most species and about 20\% for \ce{OH}. Close agreement is observed in joint probability distributions. Relative to evaluating rates at the filtered state and to a convolutional neural network baseline (requiring remeshing), reduced errors are observed and flame structure is better preserved by the GNN model. Furthermore, the model demonstrates robust generalization across varying filter widths without retraining, maintaining bounded errors and stable predictions when evaluated at coarser spatial resolutions. Additionally, in a backward facing step configuration, good accuracy is observed when assessed with species that span the entire flame. By learning on the native mesh, interpolation artifacts are avoided and spatial context important for subfilter chemistry is retained. This work can be readily extended for subgrid chemistry modeling on unstructured meshes. Finally, interpolation-free, graph-based learning provides a route to chemistry closure models in LES of practically relevant combustors.

\acknowledgement{CRediT authorship contribution statement} \addvspace{10pt}
\textbf{PD}: conceptualization, methodology, software, formal analysis, visualization, writing (original draft).
\textbf{MB}: conceptualization, methodology, supervision.
\textbf{KA}: conceptualization, methodology, formal analysis, supervision, writing (review and editing).

\acknowledgement{Declaration of competing interest} \addvspace{10pt}
The authors declare no competing financial interests or personal relationships.

\acknowledgement{Acknowledgments} \addvspace{10pt}
The authors gratefully acknowledge the computing time granted by the John von Neumann Institute for Computing (NIC) and provided on the supercomputer JURECA at Jülich Supercomputing Centre (JSC). PD acknowledges the support from the Prime Minister's Research Fellowship, India.  KA acknowledges funding support from the Core Research Grant (CRG/2023/007717) awarded by the Anusandhan National Research Foundation, India. {The support provided by Aswin Kumar Arumugam in proofreading is duly acknowledged.}

% -------------------------------------------------------------------- %
% -------------------------------------------------------------------- %
% -------------------------------------------------------------------- %
\footnotesize
\baselineskip 9pt

% -------------------------------------------------------------------- %
% -------------------------------------------------------------------- %
% -------------------------------------------------------------------- %
\clearpage
\thispagestyle{empty}
\bibliographystyle{journal}
\bibliography{main}

% -------------------------------------------------------------------- %
% -------------------------------------------------------------------- %
% -------------------------------------------------------------------- %

\newpage

\small
\baselineskip 10pt

% -------------------------------------------------------------------- %
% -------------------------------------------------------------------- %
% -------------------------------------------------------------------- %

\end{document}